# Growth and Evaluation of Improved CsI:Tl and NaI:Tl Scintillators

R. Hawrami[1*], A. Farsoni[2], H. Sabet[3], D. Szydel[4]

[1] Xtallized Intelligence, Inc., Nashville, TN 37211
[2] Avicenna Instruments, LLC, Laguna Niguel, CA 92677
[3] Massachusetts General Hospital – Harvard Medical School, Charlestown, MA 02129
[4] Deepwater Chemicals, Inc., Woodward, OK 73801

**Abstract**

Scintillators play an important role in radiation detection and imaging. Thallium-doped cesium iodide (CsI:Tl) and sodium iodide (NaI:Tl) are two of the major scintillators that have been used for many applications for many decades. In this paper we will present an improved scintillation performance of Bridgman method grown CsI(Tl) and NaI(Tl) crystals developed by Xtallized Intelligence, Inc. (XI, Inc.) and we will compare the performance with commercially available CsI:Tl and NaI:Tl. In a preliminary testing using MicroFJ−60035−TSV silicon photomultipliers (ON Semiconductor), the newly developed and improved 12.5×12.5×6 mm$^3$ CsI:Tl crystal has shown an energy resolution of 4.8% (FWHM) at 662 keV, compared to 7.2% obtained by a commercially available CsI:Tl with a size of 12.5×12.5×25 mm$^3$. Energy resolution of 5.4% (FWHM) at 662 keV is obtained for the newly improved NaI:Tl crystal, compared to 7% obtained by a commercially available 1″× 1″ NaI(Tl). Comparison of the photo peak channel locations shows that the improved CsI:Tl and NaI:Tl developed by XI, Inc. have produced much larger signal amplitudes when compared to the commercially available CsI(Tl) and NaI:Tl. Improved decay time constants are presented as well.

**Index Terms**
Gamma-ray detector, PMT, Scintillation detector, SiPM, Thallium-doped cesium iodide and sodium iodide.

## I. Introduction

As a main class of radiation detectors, scintillators play an important role in radiation detection and imaging. Alkali halides such as thallium (Tl)-doped NaI [1] and CsI [2] were among the early inorganic scintillators studied and developed. Owed to decades-long use and research into these materials they have remained widely used in various applications, such as in oil well logging [3], medical imaging [4-6], high energy/space physics [7, 8], and homeland security [9, 10]. CsI:Tl is one of the major scintillators that have been used for many applications for many decades. With a high light yield of 54,000 photons/MeV, only slight sensitivity to moisture (i.e., almost non-hygroscopic), and high atomic numbers, CsI:Tl is a reliable and efficient gamma-ray detector with an emission spectrum peaks around 540 nm. Energy resolution around 6% at 662 keV is usually achievable for CsI:Tl coupled to a photomultiplier (PMT) [11], while an energy resolution as low as 4.42% was measured when CsI:Tl was coupled to an avalanche photodiode (APD) [12]. Using digital processing, CsI:Tl coupled with SiPM has also shown an energy resolution of 5.9% at 662 keV [13]. The type of reflecting materials used to optimize light collection also affects the energy resolution [12]. NaI:Tl as a luminescent material was studied by R. Hofstadter in 1948 [1]. It has a light yield of 40,000 photons/MeV and a moderate primary decay time constant of 250 ns. Its emission spectrum peaks around 415 nm, which can be matched by the spectral response of many photomultiplier tubes. Compared to CsI:Tl, NaI:Tl is moderately hygroscopic and thus needs to be hermetically encapsulated for stability. Both NaI and CsI have cubic crystal structures, with NaI having simple cubic structure belonging in Fm-3m space group and CsI having body centered cubic structure in Pm-3m space group. Both crystals are usually grown by melt, using methods such as Bridgman and Kyropoulos. In this paper we report on modified Bridgman growth of CsI:Tl and NaI:Tl crystals with improved scintillation performance including energy resolution and light yield. Both crystals were grown by Xtallized Intelligence, Inc. (XI, Inc.) and characterized at XI, Inc., Avicenna Instruments, LLC., and Massachusetts General Hospital, Harvard Medical School.

---

* Corresponding Author. E-mail: hawrami@xtalintel.com. Phone: 1-615-916-6666.

## II. EXPERIMENTAL METHODS

High quality and high purity CsI (99.999 %) and NaI (99.999 %) raw materials in bead forms were obtained from Deepwater Chemicals, Inc. For growth CsI mixed with 0.005 wt-% Tl as dopant was loaded into a 2-inch diameter quartz ampoule, while NaI mixed with 0.005 wt-%Tl was loaded into a pre cleaned and baked 1.5-inch diameter quartz ampoules. The growth ampoules were sealed in high vacuum ($3.2\times10^{-4}$ Torr). The CsI:Tl growth ampoule was loaded into a three-zone vertical furnace, while the NaI:Tl growth ampoule was loaded into a two-zone vertical furnace, all growth runs employing the vertical Bridgman technique. Our best results were obtained when using a growth rate of 3 to 3.5cm/day with ~35 °C above the melting points of CsI:Tl and NaI:Tl compounds, respectively. After the growth process was completed, each crystal boule was retrieved from the ampoule and samples of different sizes were cut and shaped using a diamond wire saw, and processed (i.e., lapped and polished) for characterization. Scintillation performance benchmarks like energy resolution, relative light yield, non-proportionality, decay time, and afterglow characteristics were measured.

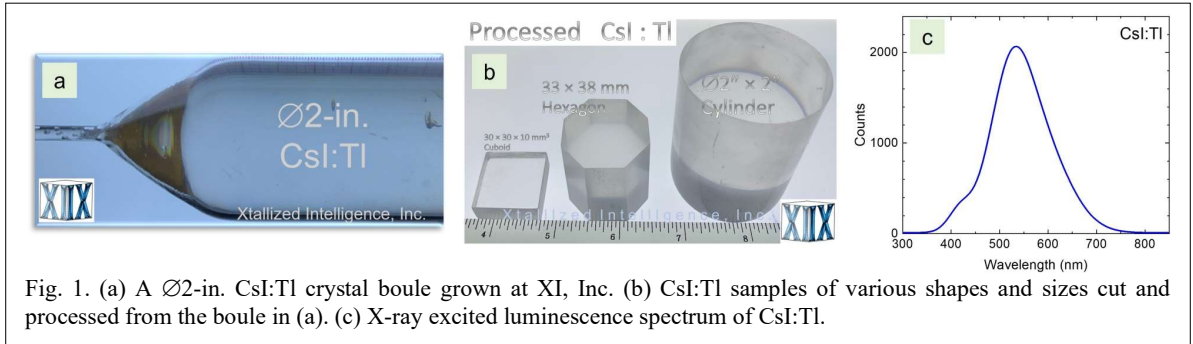

Fig. 1. (a) A ⌀2-in. CsI:Tl crystal boule grown at XI, Inc. (b) CsI:Tl samples of various shapes and sizes cut and processed from the boule in (a). (c) X-ray excited luminescence spectrum of CsI:Tl.

A $12.5\times12.5\times6$ mm$^3$ CsI:Tl sample cut from the grown boule was characterized using both SiPM and PMT. Its energy resolution and light yield were compared to a commercially available CsI:Tl crystal (from Hilger Crystals) with a size of $12.5\times12.5\times25$ mm$^3$. A NaI:Tl sample with a size of 1 in$^3$ was characterized and compared to a commercially available 1″ × 1″ NaI(Tl) (from St. Gobain Crystals). The CsI:Tl crystal was evaluated using a 2×2 array of MicroFJ−60035−TSV SiPMs, which consist of 6.07 mm active area and 6.13 mm packaging dimension. A custom-made enclosure was designed and built using a 3-D printer to accommodate the crystal and SiPMs while sealing the detector against ambient light. Each CsI(Tl) crystal was wrapped on 5 sides using a layer of ESR (3M™ Enhanced Specular Reflector) to maximize the light collection before inserted into the enclosure. The crystal was then coupled to the SiPM array using optical grease. Signals from the detectors were processed using custom-built analog electronics (with 8 μs shaping time). The CsI:Tl sample grown at XI, Inc. were tested in-situ prior to characterization at Avicenna Instruments, LLC. Another CsI:Tl sample, from a crystal grown by XI, Inc., was sent to MGH Lab for afterglow characterization.

For crystal sample characterization using a PMT at XI, Inc., a R6231-100 51 mm diameter super bi-alkali PMT from Hamamatsu, with a E1198-26 voltage divider, a Canberra 2005 pre-amplifier, and a Canberra 2022 spectroscopy amplifier were utilized. Amplifier unipolar output signals were then analyzed with a MCA8000D Amptek multichannel analyzer to generate differential pulse height spectra. γ-ray spectra from various check sources were collected by each crystal sample and the collected data were used to calculate the energy resolution, relative light yield, and non-proportionality information. Output signals from the PMT's anode were collected with a CAEN DT5720 digitizer and analyzed offline to calculate the decay time constants of each sample. The X-ray excited luminescence spectrum was performed with an X-ray tube source providing <30 keV x-rays and the spectral counts were collected by a fiberoptic-coupled Ocean Optics spectrometer, employing a silicon CCD readout.

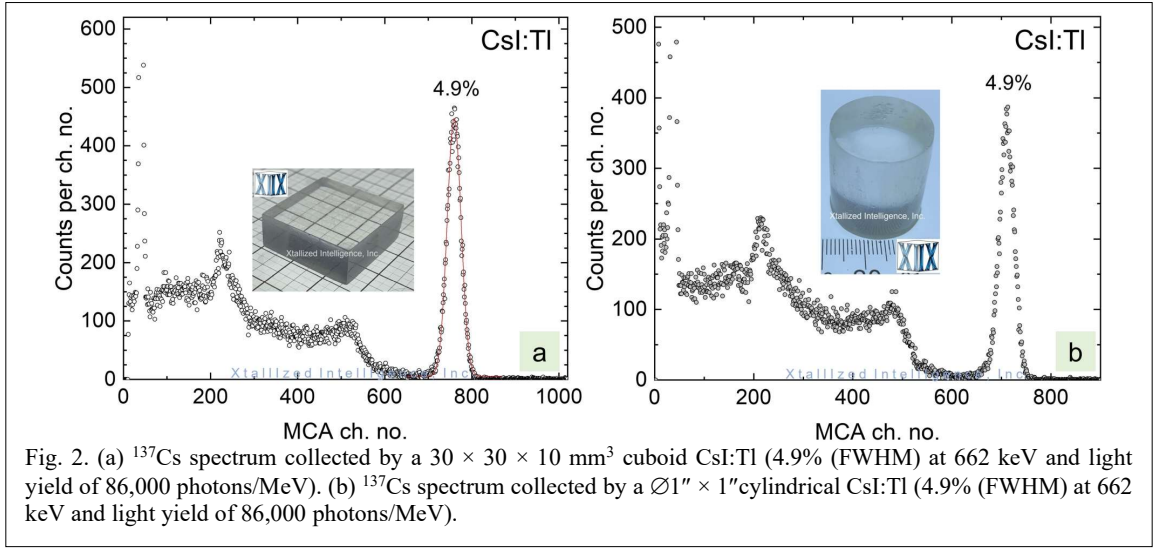

Fig. 2. (a) $^{137}$Cs spectrum collected by a 30 × 30 × 10 mm$^3$ cuboid CsI:Tl (4.9% (FWHM) at 662 keV and light yield of 86,000 photons/MeV). (b) $^{137}$Cs spectrum collected by a ∅1″ × 1″ cylindrical CsI:Tl (4.9% (FWHM) at 662 keV and light yield of 86,000 photons/MeV).

At the MGH Lab the decay time measurement was conducted by utilizing Tektronix MSO64, a 4GHz oscilloscope with 25G samples/second to record the signals. The crystal was coupled to a Hamamatsu R6237 PMT biased at 1000V and exposed to a $^{137}$Cs point source. Afterglow procedure was performed by exposing the CsI:Tl-PMT detector to X-ray source at 30 kVp with pulse duration of 1 us. A total of 100 pulses per measurement were acquired and processed up to 10 ms postexposure. For comparison, we also measured the afterglow response of a 10x10x5 mm3 CsI:Tl crystal (product of Hilger Crystals).

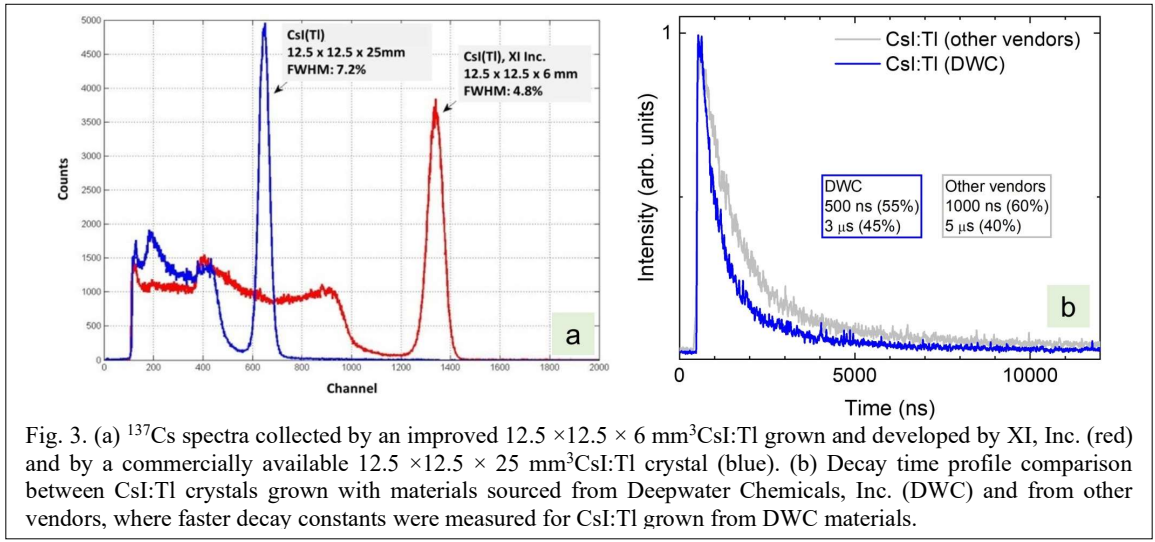

Fig. 3. (a) $^{137}$Cs spectra collected by an improved 12.5 ×12.5 × 6 mm$^3$CsI:Tl grown and developed by XI, Inc. (red) and by a commercially available 12.5 ×12.5 × 25 mm$^3$CsI:Tl crystal (blue). (b) Decay time profile comparison between CsI:Tl crystals grown with materials sourced from Deepwater Chemicals, Inc. (DWC) and from other vendors, where faster decay constants were measured for CsI:Tl grown from DWC materials.

### III. RESULTS AND ANALYSIS

*CsI:Tl*

The 2-inch diameter CsI:Tl crystal boule grown at XI, Inc. is presented in Fig. 1(a), with a few samples of various shapes and sizes from the boule shown in Fig. 1(b). The x-ray exited luminescence spectrum of CsI:Tl with an emission peak at 543 nm is shown in Fig. 1(c).

$^{137}$Cs spectra from several samples are shown in Fig. 2. Energy resolution (FWHM) of 4.9% at the full energy peak of 662 keV was calculated from the $^{137}$Cs spectrum collected by a 30×30×10 mm$^3$ cuboid (Fig. 2(a)). Light yield of 87,000 photons/MeV was estimated from comparison with a ∅1″× 1″ NaI:Tl (light yield of 42,000 photons/MeV). With a ∅1″×1″cylindrical CsI:Tl (Fig. 2(b)), energy resolution (FWHM) of 4.9% at 662 keV and light yield of 86,000 photons/MeV were calculated.

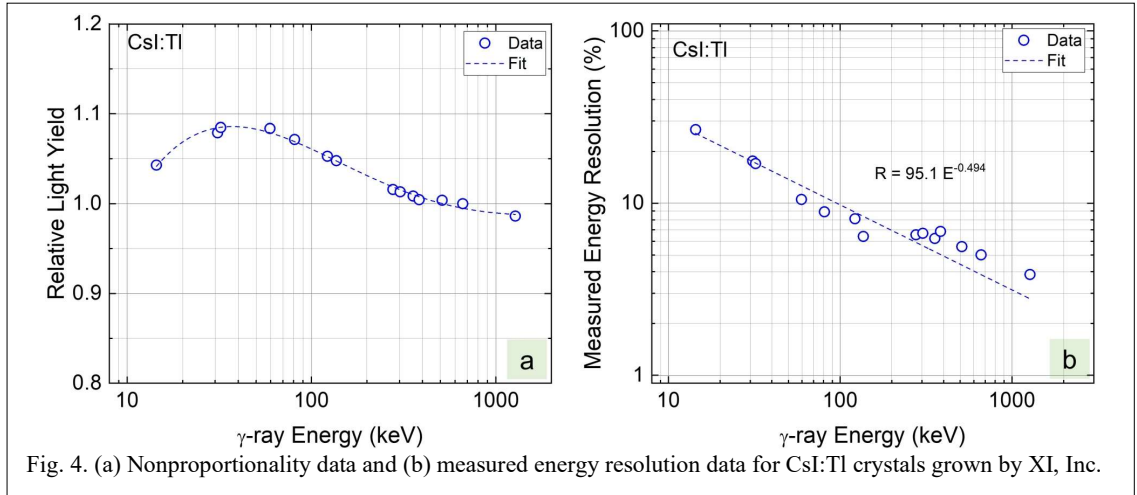
Fig. 4. (a) Nonproportionality data and (b) measured energy resolution data for CsI:Tl crystals grown by XI, Inc.

A 12.5×12.5×6 mm$^3$ CsI:Tl sample cut from the grown boule was sent to Avicenna Instruments, LLC. The crystal sample was characterized using an array of 2 × 2 MicroFJ−60035−TSV SiPMs and compared to a commercially available CsI:Tl crystal with a size of 12.5 × 12.5 × 25 mm$^3$ obtained from Hilger Crystals. Comparison of $^{137}$Cs spectra collected by either crystal is shown in Fig. 3(a), with energy resolutions (FWHM) of 4.8% for the CsI:Tl crystal grown by XI, Inc. and 7.2% for the commercially available CsI:Tl. Comparison of the photopeak channel locations shows that the improved CsI:Tl developed by XI, Inc. has produced much larger signal amplitudes (roughly double) when compared to the commercially available CsI:Tl.

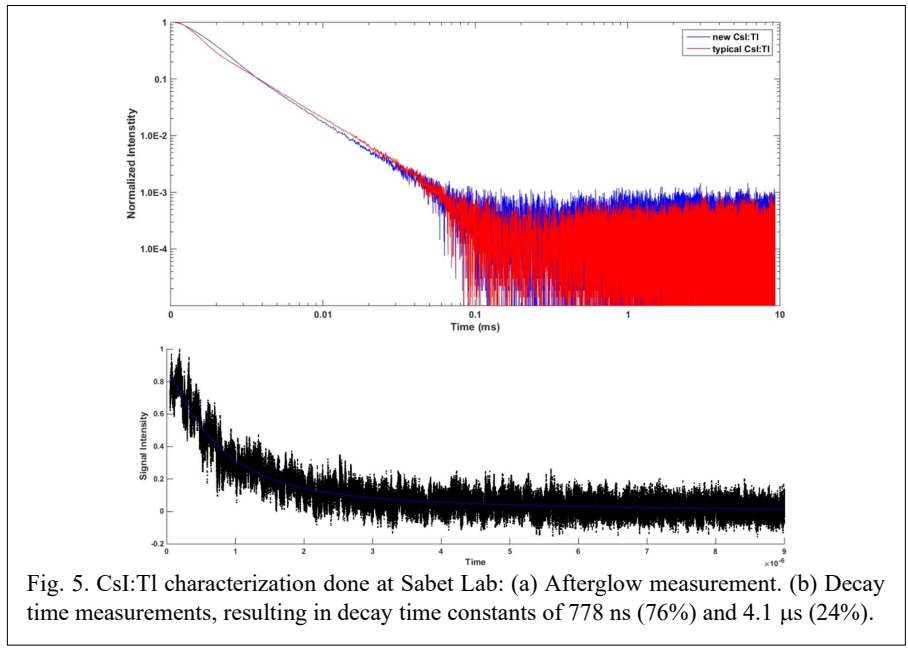
Fig. 5. CsI:Tl characterization done at Sabet Lab: (a) Afterglow measurement. (b) Decay time measurements, resulting in decay time constants of 778 ns (76%) and 4.1 μs (24%).

XI, Inc. obtained CsI:Tl starting materials from several sources, including recently from Deepwater Chemicals, Inc. Fig. 3(b) shows the decay time profile comparison between CsI:Tl crystals grown with materials sourced from Deepwater Chemicals, Inc. (DWC) and from other vendors, where faster decay constants were measured for CsI:Tl grown from DWC materials (500 ns (55%), 3 μs (45%)).

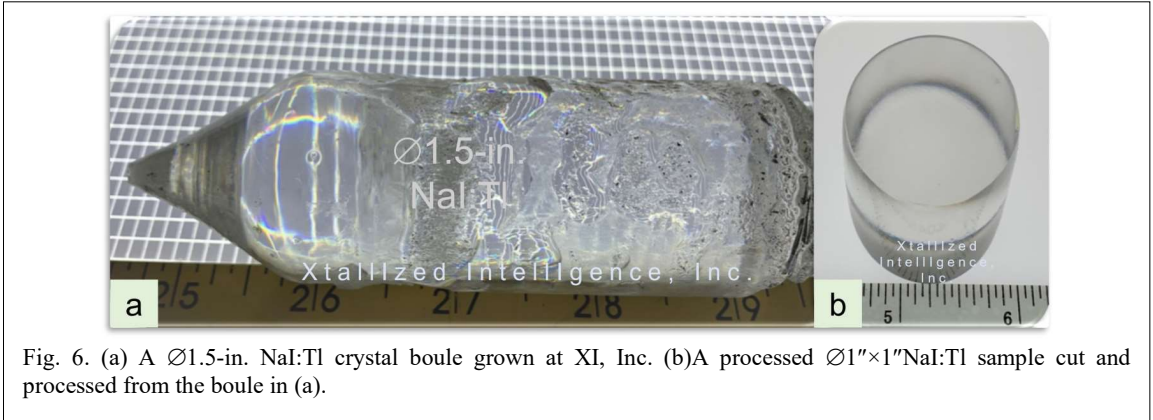

Fig. 6. (a) A ⌀1.5-in. NaI:Tl crystal boule grown at XI, Inc. (b) A processed ⌀1″×1″ NaI:Tl sample cut and processed from the boule in (a).

Nonproportionality (relative light yield as a function of energy) and measured energy resolution data, both calculated from spectra collected with a Hamamatsu R6231-100 PMT, are shown in Fig. 4. The nonproportionality data is comparable to nonproportionality behavior of typical CsI:Tl crystals. The measured energy resolution data was proportional to $0.95\ E^{-0.494}$, which is close to the expected $1/\sqrt{E}$ behavior.

The afterglow characterization at MGH shows that the new CsI:Tl grown by XI, Inc. has a similar afterglow when compared with a conventional crystal. Experimental results at MGH also show better decay constants compared with a conventional/typical CsI:Tl crystal. A rise time at 44 ns as well as primary decay time constant at 778 ns (76%) and long decay time constant at 4.1 μs (24%) were measured.

### *NaI:Tl*

The 1.5-inch diameter NaI:Tl crystal boule grown at XI, Inc. is presented in Fig. 6(a), with a processed ⌀1″×1″ sample cut from the boule shown in Fig. 6(b). A sample cut from the boule and processed characterized by radiation measurement. Fig. 7(a) shows the $^{137}$Cs spectrum collected by the crystal sample (inset), with a calculated energy resolution (FWHM) of 5.4% at 662 keV. Comparison with a commercially available ⌀1″×1″ NaI:Tl crystal (Fig. 7(b)) indicates that the light yield of the new NaI:Tl crystal is approximately 120% of the light yield of the commercially available NaI:Tl.

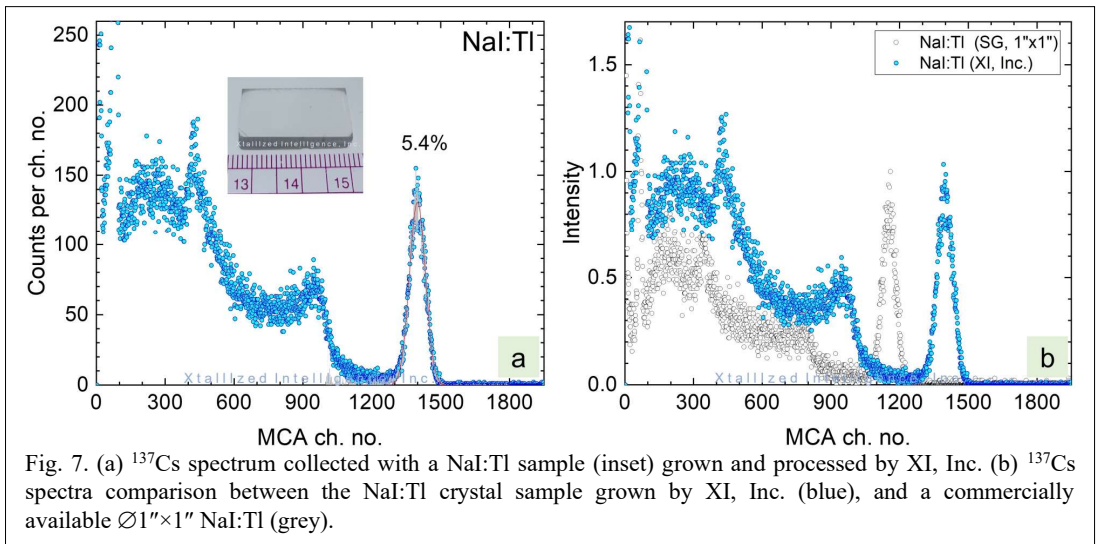

Fig. 7. (a) $^{137}$Cs spectrum collected with a NaI:Tl sample (inset) grown and processed by XI, Inc. (b) $^{137}$Cs spectra comparison between the NaI:Tl crystal sample grown by XI, Inc. (blue), and a commercially available ⌀1″×1″ NaI:Tl (grey).

Fig. 8(a) shows the nonproportionality data for NaI:Tl grown by XI, Inc. The relative light yield as a function of photon energy for this NaI:Tl is similar to the nonproportionality behavior of typical NaI:Tl crystals. The time profile for NaI:Tl is shown in Fig. 8(b), with a calculated primary fast decay constant of 234 ns (93%) and a long decay constant of 2.6 μs (7%)

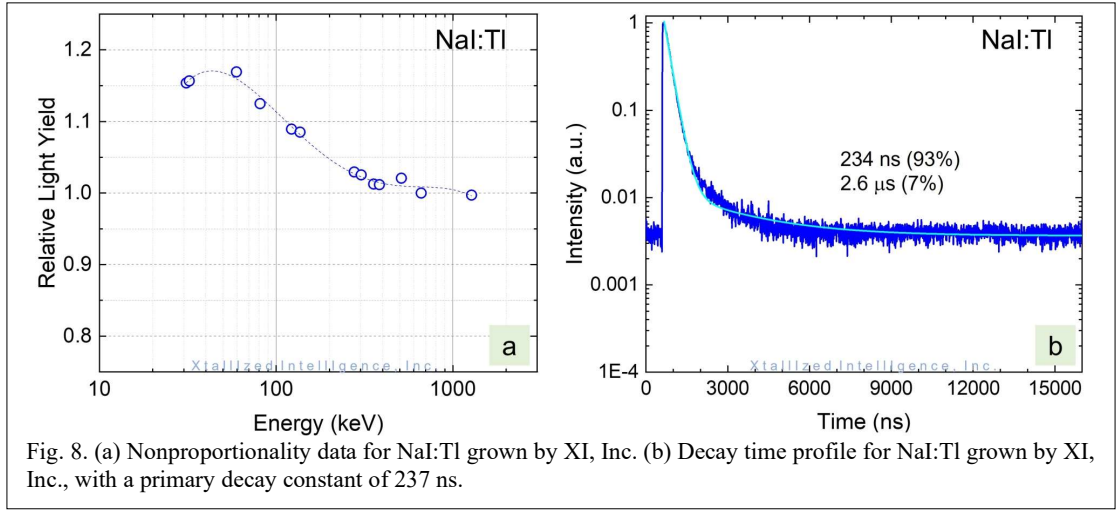

Fig. 8. (a) Nonproportionality data for NaI:Tl grown by XI, Inc. (b) Decay time profile for NaI:Tl grown by XI, Inc., with a primary decay constant of 237 ns.

### IV. CONCLUSIONS

This paper presents an improved scintillation performance of a Bridgman method-grown CsI:Tl and NaI:Tl crystals developed by XI, Inc. and performance comparison with commercially available CsI:Tl and NaI:Tl. Coupled with MicroFJ−60035−TSV silicon photomultipliers (ON Semiconductor), the newly developed and improved 12.5×12.5×6 mm$^3$ CsI:Tl cuboid crystal produced by XI, Inc. showed an energy resolution of 4.8% (FWHM) at 662 keV, compared to 7.2% obtained by a commercially available CsI:Tl with a size of 12.5×12.5×25 mm$^3$. Similar performance (energy resolution of 4.9%) was obtained for when XI, Inc.-produced CsI:Tl cuboid was coupled to a R6231-100 photomultiplier tube (Hamamatsu). Decay time constants for CsI:Tl were improved, while afterglow was similar to other commercially available CsI:Tl crystals. Growth of NaI:Tl with an improved performance was also presented. Energy resolution of 5.4% at 662 keV was measured with a photomultiplier tube. Light yield of the new NaI:Tl crystal was approximately 20% higher than other commercially available NaI:Tl crystals.

### V. ACKNOWLEDGMENTS

The work in this paper was supported in parts by Deepwater Chemicals, Inc. The authors would like to thank Drs. Winston Chern and Rajiv Gupta for afterglow measurements.